%% file: 5GV2X_useCase_reqs_Final.tex
\documentclass[twocolumn, 10pt, journal]{IEEEtran}
\usepackage{color}
\usepackage{setspace}
\usepackage[cmex10]{amsmath}
\usepackage{psfrag}
\usepackage{subfigure}
\usepackage{hyperref}
\usepackage{multirow}
\usepackage{amsmath}
\usepackage{multicol}
\usepackage{setspace}
\usepackage{amsmath}
\usepackage{subfigure} 
\usepackage{multicol}
\usepackage{algorithmic}
\usepackage{algorithm}
\usepackage{epsfig}
\usepackage{cite}
\usepackage{graphicx} 
\usepackage{geometry}
\usepackage{color}
\usepackage{xcolor}

\newcommand*{\savedfootnotes}{}
\newcommand*{\resetsavedfootnotes}{\global\let\savedfootnotes\empty}
\newcommand{\tablefootnote}[1]%
  {%
    \footnotemark
    \xdef\savedfootnotes%
      {\unexpanded\expandafter{\savedfootnotes}\noexpand\footnotetext{#1}}%
  }
\edef\endtable%
  {%
    \aftergroup\noexpand\savedfootnotes
    \aftergroup\noexpand\resetsavedfootnotes
    \unexpanded\expandafter{\endtable}%
  }
\usepackage{tikz}
\def\checkmark{\tikz\fill[scale=0.4](0,.35) -- (.25,0) -- (1,.7) -- (.25,.15) -- cycle;}

\begin{document}
\date{}
\title{%
 Use Cases, Requirements, and Design Considerations for 5G V2X
%\Large{A Vision Towards Collaborative, Efficient, End-to-End, Autonomous Mobility} 
} 
\author{
  		Mate Boban, Apostolos Kousaridas, Konstantinos Manolakis, %et al. \\%
  		Joseph Eichinger, Wen Xu\\
        Huawei Technologies, German Research Center, 80992 Munich, Germany \\
	Email: mate.boban@huawei.com \\
}

\maketitle

\input{abstract}

\input{intro}

\input{useCasesV3}

\input{commSystems}

\input{conclusions}

\input{ack}

\bibliographystyle{IEEEtran}

\bibliography{draftIII_tex}
\clearpage 

\end{document}

%% file: abstract.tex
\begin{abstract}

%Future communication systems will need to employ a range of complementary technologies to support a diverse set of vehicle-to-everything (V2X) use cases, ultimately leading to accident-free, cooperative autonomous vehicles that use the available roadway efficiently. 

Ultimate goal of next generation Vehicle-to-everything (V2X) communication systems is enabling accident-free cooperative automated driving that uses the available roadway efficiently. To achieve this goal, the communication system will need to enable a diverse set of use cases, each with a specific set of requirements.
%employ a range of complementary technologies to support a diverse set of V2X use cases. 
We discuss the main use case categories, analyze their requirements, and compare them against the capabilities of currently available communication technologies. Based on the analysis, we identify a gap and point out towards possible system design for 5G V2X that could close the gap. Furthermore, we discuss %We describe 
an architecture of the 5G V2X radio access network that incorporates diverse communication technologies, including current and cellular systems in centimeter wave %in the ``Day One'' 5G deployment (year 2020), as well as 
and millimeter wave, IEEE 802.11p and vehicular visible light 
communications. % beyond Day One. 
%We discuss the suitability of each technology to support different V2X use cases, with more complex use cases requiring inter-operation of several technologies. 
%We identify the enhancements required to existing communications systems to enable high reliability, low latency services, ranging from physical layer technologies (synchronization, frame structure, numerology, multiple antenna systems) to medium access (scheduling), to cross layer and architectural aspects (mobile edge computing, slicing, connection management). 
Finally, we discuss the role of future 5G V2X systems in enabling more efficient vehicular transportation:  %(reducing the per-vehicle parking area in half), 
from improved traffic flow through reduced inter-vehicle spacing on highways and coordinated intersections in cities (the cheapest way to increasing the road capacity), to automated smart parking (no more visits to the parking!), ultimately enabling seamless end-to-end personal mobility.

\end{abstract}

%% file: intro.tex
\section{Introduction} \label{sec:Intro}
%\footnote{\mate{Mate's comments}, \konstantinos{Konstantinos's comments}, \chan{Chan's (RAN) comments}}
Personal mobility and vehicular transportation systems in general are undergoing somewhat of a revolution. The reasons for this can be found in the new societal and market trends. The main new societal trends affecting the transportation are: i) new wave of urbanization creating pressure on the existing transportation infrastructure, which cannot grow as fast as the demand; ii) ever more stringent emission- and energy-related regulation; and iii) high pressure on public transport and logistics/delivery services to become more adaptive and dynamic. The key market trends are: i) the advent of automated driving; ii) new modes of car use and  ownership (i.e., a shift towards the ``shared economy''); and iii) live and open data availability, including crowd sourcing and open platforms, which enables more efficient use of transportation resources. %; and iii) . % a on the market by year 2020. 
These trends are creating a shift towards more reactive and intelligent transport infrastructure, with the following goals:
\begin{itemize}
\item Accident-free transportation;
\item Supporting higher traffic flow (i.e., increasing the road surface utilization n-fold, currently standing below 10~\%~\cite{shladover2009cooperative});
\item Higher vehicle utilization (e.g., increasing the average personal car utilization well above the current 5\%); % to 50\%)
\item More efficient/greener transport (zero emission vehicles).
\end{itemize}
%We believe that 

Communication technologies, in the form of Vehicle-to-everything (V2X) communication%\footnote{The term Vehicle-to-everything (V2X) subsumes all of the following communication link types: %A vehicle can make use of different links to communicate with other machines and operate Intelligent Transport System (ITS) services~\cite{seo2016lte}: 
%a) Vehicle-to-Vehicle (V2V); b) Vehicle-to-Infrastructure (V2I) (e.g., traffic lights); c) Vehicle-to-Pedestrian (V2P); and d) Vehicle-to-Network (V2N), where the vehicle connects to a backend server (e.g., OEM server) in the network to enable additional services like map updates, fleet-based data collection, etc. 
%}
\footnote{In terms of the naming conventions, we consider that the term Vehicle-to-everything (V2X) subsumes all of the following communication modes: a) Vehicle-to-Vehicle (V2V); b) Vehicle-to-Infrastructure (V2I) (e.g., communication with roadside units (RSUs), traffic lights, or, in case of cellular network, base station); c) Vehicle-to-Pedestrian (V2P); and d) Vehicle-to-Network (V2N), where the vehicle connects to an entity in the network (e.g., a backend server or a traffic information system). %to enable additional services like map updates, fleet-based data collection, etc.
}, will play a key role in reaching these goals. While the in-vehicle sensors can enable many functionalities without the use of inter-vehicle communication, %V2X 
%Specifically, 
the benefits that V2X communication %can bring  %such mobility paradigm 
%are multiple:
%\begin{itemize}
%\item The first 
promises are: i) safer driving as a consequence of enabling the well-studied safety use cases~\cite{etsi14,3gppTR22886}; and % Eventually, through the combination of cooperation and automated driving, V2X will help achieve accident free driving (or, at minimum, eradicate the accidents caused by human driver errors).
%\item V2X can 
ii) improved road capacity due to better road~\cite{shoup2006cruising} and parking infrastructure~\cite{ferreira2014self} utilization. %, % and system-level traffic coordination, 
The goals of this paper are to: 
\begin{itemize}
\item consolidate and analyze relevant use cases and their requirements that will drive the 5G V2X communication system design;
\item discuss the capabilities of existing communication technologies to support the 5G V2X use cases, including the resulting gap analysis; 
\item provide guidelines on how existing and future communication systems can be leveraged as part of a 5G V2X solution.
\end{itemize}
As a starting point, below we elaborate on the potential benefits of V2X communication through an illustrative example that involves many of the use cases described in~\cite{etsi14} and~\cite{3gppTR22886}.

%\iffalse
\begin{figure*}[!t]
  \begin{center}
	\includegraphics[trim=0cm 1.9cm 0cm 0.1cm,clip=true,width=0.85\textwidth]{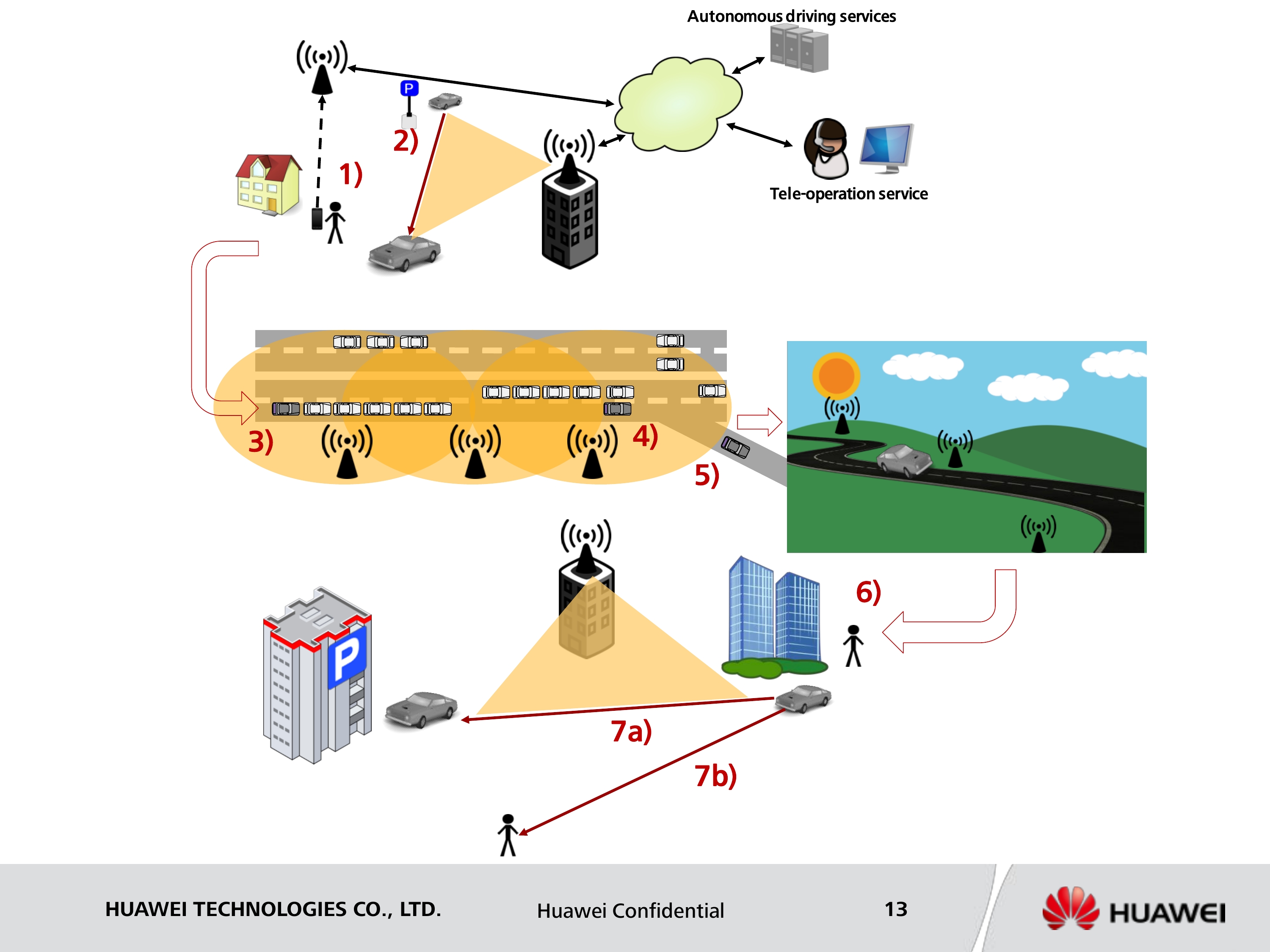}
  \end{center}
\caption{Motivating example for 5G V2X. To enable the described scenario, the following will be needed: i) incorporating existing use cases (e.g., those found in~\cite{etsi14} and~\cite{3gppTR22886} and new use cases; and ii) extending the existing network technologies so they can support reliable, low-latency, ubiquitous nature of the involved use cases.}
%Communication technologies.}
\label{fig:MotivatingExample}
\end{figure*}

%\subsection*{Motivating Example:} \label{sec:Motivation}
Figure~\ref{fig:MotivatingExample} shows %a motivating 
an aspirational example of what future personal mobility might look like, exemplified through several 
salient use cases, such as those found in~\cite{etsi14} and ~\cite{3gppTR22886}, some of which are shown in Fig.~\ref{fig:FutureUseCases}. %for what future connected vehicles might achieve. 
The following use cases are employed in consecutive steps and denoted in Fig.~\ref{fig:MotivatingExample} accordingly). %, consider the scenario in Fig.~\ref{figgy}. The example shows the 
\begin{enumerate}
\item Vehicle-on-demand: a vehicle user\footnote{Note that we use the term ``vehicle user'' to distinguish the new mobility models from conventional ``driver'' and ``passenger'' models.} hails a car via an app depending on purpose (e.g., on a weekday: family sedan; on a sunny weekend: convertible).
\item The vehicle self-drives or is tele-operated~\cite{hosseini2014interactive} (either by human or remote robot/artificial intelligence) to the user. User's personal transportation app connected over the mobile radio network adjusts the car to his presets.
%\begin{itemize}
%\item Adjusts the car to his presets;
%\item Loads music/videos from cloud where he left of at home;
%\item Pays for tolls.
%\end{itemize}
\item The vehicle is capable to drive itself: it searches for the platoon best suited for the vehicle user's preferences and performs cooperative maneuvering %can be added on the highway scenario because the car is looking to the best platoon
%it performs lane changing maneuver and joins 
to join selected platoon, thus saving energy by allowing car following at very short gaps.
\item Two-way navigation (i.e., feeding and obtaining real time traffic info) guides the vehicle around a congested stretch of the road.
\item The user takes over control to exit the highway and drive by himself on a particularly scenic road suggested by the connected navigation system.
%\item Voice-activated restaurant app gets you a table
%\item Car drops you off for dinner and parks itself
\item The vehicle drops the user off at the destination. 
\item The vehicle self-drives~\cite{SchwesingerVBS15}, either to an automated parking lot~\cite{ferreira2014self} -- 7a) -- or to a different user -- 7b). Automated electric vehicles might not need parking garages in the conventional sense -- for storing the vehicle while not in use. Rather, as noted in~\cite{Economist17}, ``garages might become service centres where shared battery-powered cars could be cleaned, repaired and recharged before being sent back on the road.''
\end{enumerate}

While this example might seem far into the future, most of the technology needed to enable it (high precision maps, real time traffic information, sensors inside the vehicle such as radars, cameras, ultrasonic, etc.) are either already available or will be in the near future. %As the motivating example in Section~\ref{sec:Motivation} describes, 
The most prominent missing component is high reliability, low latency communications system. Convergence of communication technologies with advanced sensors inside the vehicle, combined with ubiquitous network connectivity and available traffic information data, can give rise to seamless mobility between user's origin and destination, without the need to pick up, drop off, or even own a vehicle.
The benefits of such mobility paradigm are multiple:
\begin{itemize}
\item Due to better road utilization and system-level traffic coordination, the road capacity is improved, having as a consequence reduced congestion, better fuel economy, and less time spent in traffic.
\item The user goes directly from his doorstep to the destination, without needing to go to through the intermediate steps of: i) going to pick up the car in a parking; ii) driving to another parking instead of the destination; iii) walking from the parking to the destination.
\item The on-demand vehicle also increases the vehicle utilization -- instead of being idle while the single owner is not using it, the vehicle is being used by multiple users (akin to existing car-sharing services DriveNow and Car2Go owned by BMW and Daimler, respectively).
\item Since vehicles are utilized more efficiently (i.e., the idle time is reduced) and combined with automated smart parking lots~\cite{ferreira2014self}, the number of parking spots required in the system can be decreased considerably. This brings a clear economic and ecological benefit, since parking lots account for the single biggest land mass use in large developed cities~\cite{shoup2006cruising}. %\cite{shoup2005high}.
\item Some estimates in large cities point out that up to 30\% of traffic on the streets is generated by drivers searching for parking (i.e., vehicles that already arrived at the destination)~\cite{shoup2006cruising}; the new mobility paradigm has the potential to completely remove the additional road traffic generated by searching for parking. %, with the reduced number of parking spots required combined with real-time knowledge of parking availability.
\item In addition to increased efficiency, through the combination of cooperation and automated driving, the new mobility paradigm would lead to accident free driving (or, at minimum, eradicate the accidents caused by human driver errors).
\end{itemize}
%\fi

Future vehicle can be perceived as a powerful mobile device that is equipped with various sensors (camera, radar, ultra-sound, etc.), having adequate computational resources to support a wide range of automotive services. Many current efforts regarding automated cars adopt a non-cooperative approach for automated driving, wherein the key goal is to replace human driver with a ``robot'' driver: a combination of sensors and software in the vehicle. However, while replacing a human with a robot driver might have a limited impact in terms of safety, the impact on road throughput is minimal~\cite{shladover2009cooperative}. 
On the other hand, connected vehicles (either automated, tele-operated, or human driven) have the potential to cooperate  in order to improve traffic flow on highways \cite{shladover2009cooperative}, in intersections~\cite{dresner2008multiagent}, and in parking lots~\cite{ferreira2014self}, increase safety (by “seeing” around the corner – capability unparalleled by other sensors) and reduce energy consumption \cite{shladover2009cooperative}. 
%Interconnected vehicles %connected to the Internet 
%will also %communicate directly with other machines 
%be able to extend the perception of one vehicle beyond the capabilities of its integrated sensors’ range, to assist in exchanging vehicle related information (e.g., cooperative awareness, road hazards, etc.) and to improve decision making for purposes of safer (self-)driving. %Some examples that indicate the benefits of communication technologies (e.g., IEEE 802.11p, 3GPP LTE, and 5G) for road traffic are illustrated in Fig. 2.
%A vehicle can make use of different links to communicate with other machines and operate Intelligent Transport System (ITS) services~\cite{seo2016lte}: a) Vehicle-to-Vehicle (V2V); b) Vehicle-to-Infrastructure (V2I) (e.g., traffic lights); c) Vehicle-to-Pedestrian (V2P); and d) Vehicle-to-Network (V2N), where the vehicle connects to a backend server (e.g., OEM server) in the network to enable additional services like map updates, fleet-based data collection, etc. The term Vehicle-to-Anything (V2X) subsumes all of the aforementioned communication links. %describes any communication involving a vehicle as a source or destination of a message. At this point we should clarify that V2V, V2P and V2I modes could be supported either via a Base Station or via an adhoc communication.

In order to achieve the vision described above, a set of enabling use cases needs to be implemented, many of which rely heavily on the communication system connecting vehicles with other nearby vehicles and with infrastructure/backend. In this paper we analyze these use cases and their requirements and analyze to what extent the existing technologies can satisfy these requirements. Furthermore, we summarize an architecture for V2X radio access network, %an architecture for future 5G V2X systems, 
 which provides native support for the requirements that the various V2X use cases have, most notably low latency, high reliability, and scalability.
%We discuss the enabling communication technologies and how they can help support existing use cases. 
%In addition to existing use cases, we also describe a set of future use cases that will ultimately lead to end-to-end, accident-free, efficient vehicular transportation. % through the use of V2X communication.

The rest of the paper is organized as follows. Section~\ref{sec:UseCases} describes relevant V2X use cases and analyzes their requirements. % and indicates which communication technologies can enable them. 
Section~\ref{sec:CommSystems} describes communication technologies for enabling V2X. 
%Section~\ref{sec:Technologies} describes the most important V2X requirements in light of 5G network design goals, based on which 
%Section~\ref{sec:Architecture} defines components of 5G radio access network architecture needed to support V2X use cases. 
Section~\ref{sec:Gap} discusses to which extent a communication technology can support each use case, along with the resulting gap analysis. It also describes the network architecture and indicates key technologies for enabling  5G V2X.
%the gap analysis results related to the most important radio access technologies %and architecture components 
%needed to enable stringent 5G V2X use cases. %Based on the gap analysis, 
Section~\ref{sec:Conclusions} concludes the paper.

%% file: useCasesV3.tex
\section{Use Cases and Requirements}\label{sec:UseCases}

\begin{figure*}[!t]
	\begin{center}
		\subfigure[Platooning/Cooperative Adaptive Cruise Control (CACC)~\cite{3gppTR22886}.]{\label{fig:CACC}\includegraphics[trim=0cm 2cm 0cm 2.5cm,clip=true,width=0.32\textwidth]{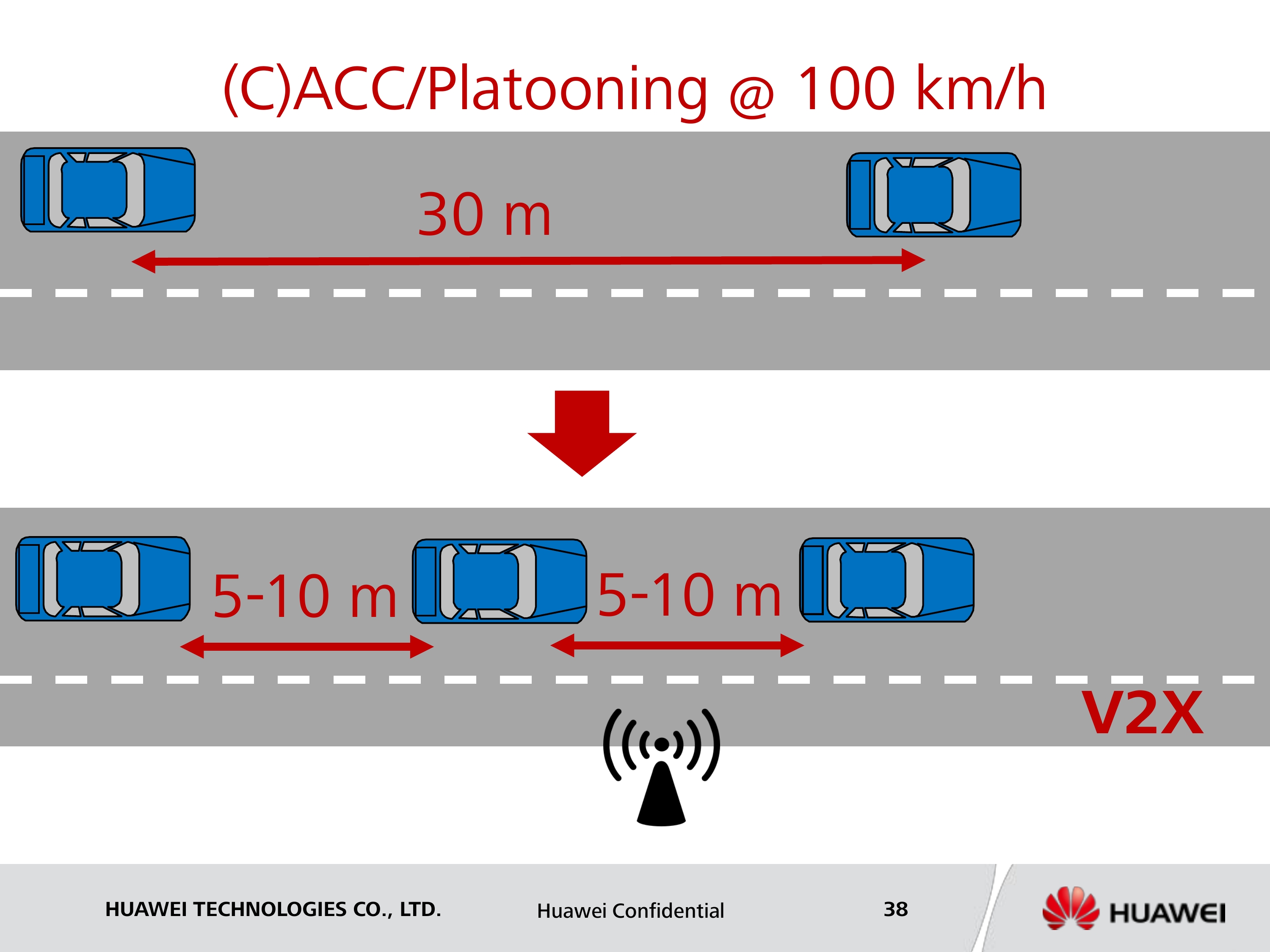}}
		\hspace{1mm}
		\subfigure[Lane (or Road) Merging~\cite{3gppTR22886}.]{\label{fig:laneMerging}\includegraphics[trim=0cm 1.8cm 0cm 2cm,clip=true,width=0.32\textwidth]{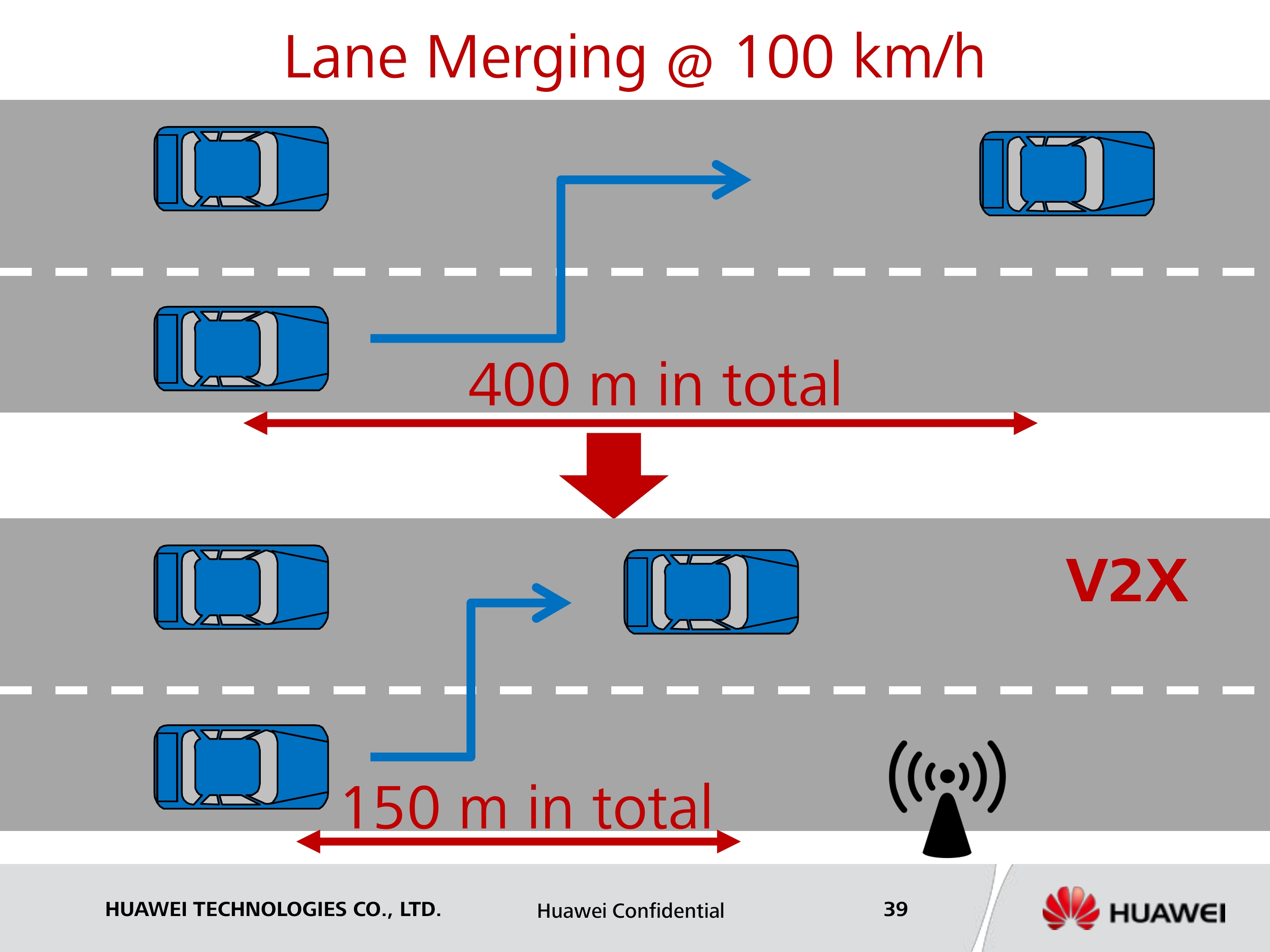}}
		\subfigure[Connected Automated Parking~\cite{ferreira2014self}.]{\label{fig:parking}\includegraphics[trim=0cm 3.5cm 0cm 0cm,clip=true,width=0.32\textwidth]{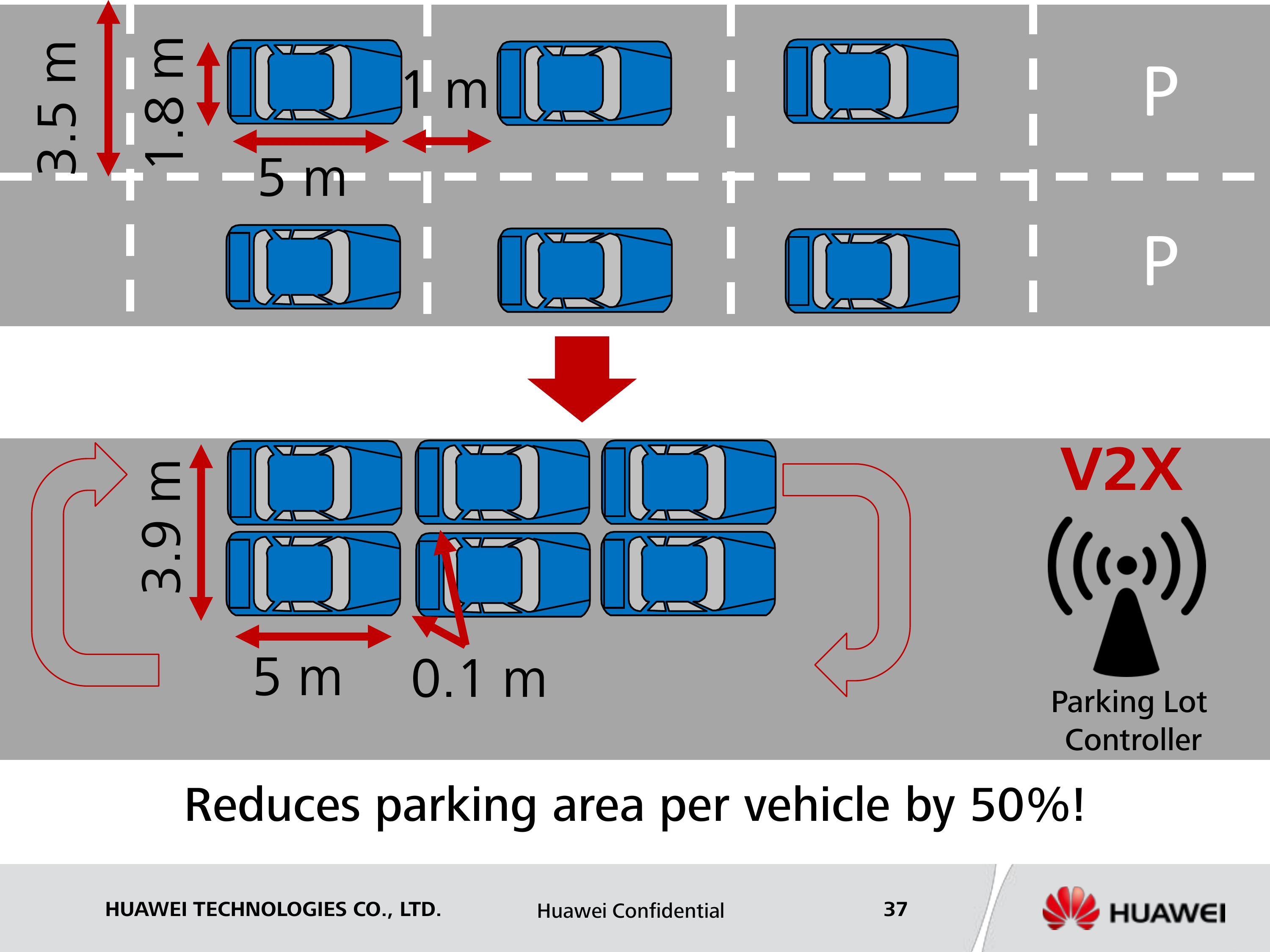}}
		\hspace{1mm}  	
		\subfigure[Cooperative Intersection Control~\cite{3gppTR22886}.]{\label{fig:Xing}\includegraphics[trim=0cm 8cm 0cm 2cm,clip=true,width=0.96\textwidth]{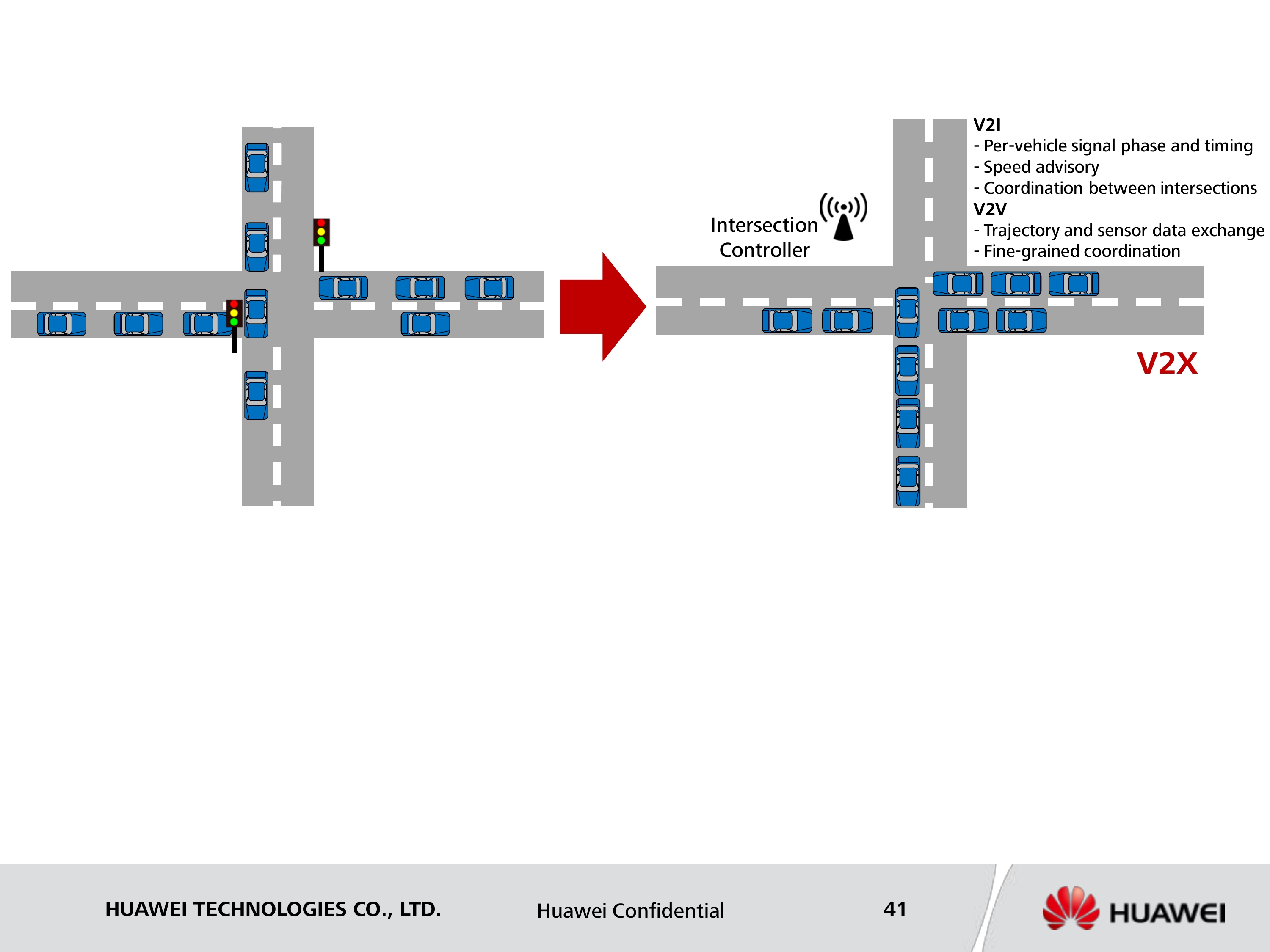}}
		\hspace{1mm}  	
	\end{center}
	\caption{Relevant advanced use cases describing the enhancements expected from 5G V2X support. The enhancements range from reduced time and space for executing maneuver, to reducing the space required to park vehicles, to increasing the traffic flow  through an intersection.}
	\label{fig:FutureUseCases}
\end{figure*}

V2X use cases focus on safety, traffic efficiency, and infotainment services. Key functional and performance requirements for safety have already been described by the European Telecommunications Standards Institute (ETSI) Intelligent Transport Systems (ITS)~\cite{etsi14}, the US Department of Transportation, and individual research projects. These use cases are used for warning and increasing the environmental awareness %of the driver and they are classified under the first level (level 1) of the taxonomy proposed by the SAE\footnote{\url{http://standards.sae.org/j3016_201401/}}. Level 1 communication is mainly 
based on periodic (e.g., Cooperative Awareness Message -- CAM~\cite{ETSIEN3026372}) or event-driven (e.g., Decentralized Environmental Notification Message -- DENM~\cite{ETSIEN3026373}) broadcast messages, with repetition rate as high as 10~Hz (e.g., emergency vehicle warning) or lower (e.g., road works warning)~\cite{etsi14}. V2X use cases have also been identified by 3GPP~\cite{3gppTR22885}, taking into account services and parameters defined in the first release of ETSI ITS~\cite{etsi14}. In this group of use cases the maximum tolerable latency is 100~ms, while the target radio layer message reception reliability is 95\%. These use cases assume a single enabling technology, namely cellular based V2X communication. Enhanced V2X (eV2X) use cases~\cite{3gppTR22885} have been defined by 3GPP as part of Release 15, including more advanced use cases such as cooperative intersection control, lane merging, and platooning (Fig.~\ref{fig:FutureUseCases}), which have more stringent requirements. 

The benefits that V2X communication brings to these use cases compared to the sensor-only solutions are multiple. In case of platooning and Cooperative Adaptive Cruise Control (CACC, Fig.~\ref{fig:CACC}) and lane merging (Fig.~\ref{fig:laneMerging}), the benefit comes from proactive communication of  locations, speeds, and trajectories, which results in shorter time and space required to execute a maneuver. In case of connected automated parking use case (Fig.~\ref{fig:parking}), communication enables centralized control and planning. Furthermore, combined with on-board sensors, it helps reduce the required lateral and longitudinal spacing between vehicles. As noted in~\cite{ferreira2014self}, this approach can halve the space needed to park a vehicle. Similarly, cooperative intersection control (Fig.~\ref{fig:Xing}) can enable more efficient intersection operation by informing the vehicles about signal phase and timing, dynamically adjusting the speed, and coordinating the flow through multiple intersections based on the current traffic conditions. Due to space limitations, we do not provide further details on each of the use cases; rather, we focus on groups of use cases and the overall requirements of those groups (Table.~\ref{tab:UseCases}). For a detailed description, we refer the reader to 3GPP technical reports~\cite{3gppTR22885,3gppTR22886} and a recent overview article~\cite{seo2016lte}.

With the increasing availability of vehicles that are capable of supporting higher automation levels, the need for coordination among vehicles and their capability to do so becomes increasingly more relevant. All automated vehicles rely on the premise that they continuously plan their trajectories and, based on the observed environment, select the driving trajectory. Due to the safety requirements, automated driving sets the most stringent performance requirements for the communication layer in terms of delay, reliability, and capacity. Cooperative Lane Change, Cooperative Collision Avoidance, Convoy Management (Platooning) are typical examples of V2X use cases that are eventually expected to lead to fully connected automated vehicles. The involved vehicles trigger a specific use case for safety reasons (e.g., emergency maneuver) or for efficient traffic flow (e.g., platooning) using the monitoring data from the installed sensors, together with the information received from neighboring vehicles. Thereinafter, the automated vehicles undertake to coordinate and plan their maneuvers or trajectories in order to address the triggering event, based on the built environmental perception. As vehicles advance towards higher automation levels and need to deal with increasingly complex road situations, there will be a need for a complementary communication technology for the exchange of cooperative information with higher bandwidth and improved reliability. In connected automated vehicles the performance requirements are more stringent, with certain use cases requiring ultra-reliable communication links ($>$99\%~\cite{alliance20155g_2}), with much lower maximum end-to-end latency (1-10 ms), and higher data rate. An initial analysis of Automated Driving use cases has recently started with the second release of ETSI ITS specification, while the conclusions of the NGMN~\cite{alliance20155g_2} and other initiatives summarized in the 5G PPP white paper on automotive sector\footnote{\url{https://5g-ppp.eu/wp.../02/5G-PPP-White-Paper-on-Automotive-Vertical-Sectors.pdf}} % ~\cite{5gppp} 
have been taken into consideration for performance requirements derivation.

%Cooperative Lane Change, Cooperative Collision Avoidance, Convoy Management (Platooning) are typical examples of V2X use cases, where connected automated vehicles participate~\cite{hobert2015enhancements}. The involved vehicles trigger a specific use case for safety reasons (e.g., emergency maneuver) or for efficient traffic flow (e.g., platooning) using the monitoring data from the installed sensors, together with the information received from neighboring vehicles. Thereinafter, the automated vehicles undertake to coordinate and plan their maneuvers or trajectories in order to address the triggering event, based on the built environmental perception. As vehicles advance towards higher automation levels and need to deal with increasingly complex road situations, there will be a need for a complementary communication technology for the exchange of cooperative information with higher bandwidth and improved reliability. In connected automated vehicles the performance requirements are more stringent, with certain use cases requiring ultra-reliable communication links (99.999\%~\cite{alliance20155g_2}), with much lower maximum end-to-end latency (1-10 ms), and higher data rate. An initial analysis of Automated Driving use cases has recently started with the second release of ETSI ITS specification, while the conclusions of the NGMN~\cite{alliance20155g_2} and other initiatives summarized in the 5G PPP white paper on automotive sector~\cite{5gppp}  have been taken into consideration for performance requirements derivation.

Vehicles will be also connected to one or more ITS application servers (e.g., for traffic management services) via V2N, which does not require strict latency or reliability requirements. However, there is a specific use case, where numerous technical hurdles need to be overcome, especially for the cellular technologies: vehicle teleoperation on public roads. Teleoperation refers to vehicles that are controlled over the network by a remote operator~\cite{hosseini2014interactive}. With the aid of sensors mounted on the vehicle, a remote operator can control one or multiple concurrent vehicles. Vehicle tele-operation could be complementary to automated driving (e.g., teleoperator taking control in particularly complex driving situations) as well as a transition technology to fully automated driving. With the promise of high reliability, availability, and sub-10 ms end-to-end delays, 5G systems have the potential to enable teleoperated vehicles. Furthermore, there are also use cases where different communication modes and technologies (V2V, V2I, sensors, ITS application sensors) could be combined e.g., Connected Automated Parking Lots, where vehicles communicate with each other and with the parking lot controller in order to park themselves into as little space as possible~\cite{ferreira2014self}.
Apart from the performance requirements, there is also a list of functional requirements that should be supported by communication technologies for the implementation of V2X use cases of connected automated vehicles. A non-exhaustive list includes: a) Different modes of dissemination: unicast, broadcast, multicast, geocast, b) Single-hop or multi-hop V2X communication range, c) Connection management, d) prioritization of V2X messages, e) Congestion control and retransmission capability.

Table~\ref{tab:UseCases} presents the main categories of V2X use cases and their key performance requirements in terms of reliability, communication latency, and the expected data rate per vehicle. % and the communication range.
%Table I presents the main categories of V2X use cases, their key performance requirements (i.e., reliability, communication latency and the expected data rate per vehicle), also highlighting the required communication mode (i.e., the available communication modes and their definitions are provided earlier in this section). 
Each identified use case type describes a specific subset of operations required by a fully cooperative and automated vehicle:
\begin{itemize}
\item Cooperative awareness: warning and increase of environmental awareness (e.g., Emergency Vehicle Warning, emergency electronic brake light~\cite{festag2014cooperative}, etc.).
\item Cooperative sensing: exchange of sensor data (e.g., raw sensor data) and object information that increase vehicles’ environmental perception.
\item Cooperative maneuver: includes use cases for the coordination of the trajectories among vehicles (e.g., lane change, platooning, CACC, and cooperative intersection control, which are shown in Fig.~\ref{fig:FutureUseCases}).
\item Vulnerable Road User (VRU): notification of pedestrians, cyclists etc
\item Traffic efficiency: update of routes and dynamic digital map update; for example, signal phase and timing (SPAT/MAP), green light optimal speed advisory (GLOSA), etc.~\cite{festag2014cooperative}.
\item Teleoperated driving: enables operation of a vehicle by a remote driver.
\end{itemize}
The latency and reliability values that are presented in Table~\ref{tab:UseCases} present the more stringent requirements as defined in the 3GPP document on V2X requirements for 5G~\cite{3gppTS22186}. Many of the use cases (e.g., cooperative maneuver) could also be supported using more lenient values, at the cost of less optimal operation (e.g., % but minimum latency allows 
in case of cooperative maneuver, larger distances and slower %shorter distances and faster 
execution of the maneuver). %which improves the driving performance.
%The minimum and/or maximum latency, reliability and data rate values are presented in the table below. 
 For some use case types the range of latency and data rate values is large, because there are various use cases under each category that have different requirements e.g., a cooperative maneuver in an emergency situation has more stringent delay and reliability requirements comparing to a normal lane change. The reliability is directly influenced by the required latency; the lower the end-to-end latency requirement of a transmission, the higher the expected reliability. It should be also noted that some use cases (e.g., platooning) could be supported using higher latency values, but the minimum latency allows shorter distances and faster execution of the maneuver, which improves the driving performance.

%\begin{table}[t!]
%\centering
%\caption{V2X Use Cases and Performance Requirements} 
%\label{tab:UseCases}
%
%\begin{tabular}{l c c c c}
%\textbf{Use Case Type}	& \textbf{V2X Mode}	& \textbf{End-to-End Latency}	& \textbf{Reliability}	& \textbf{Data Rate per vehicle (kbits/sec)}\\
%Cooperative Awareness\footnote{In the ``Cooperative Awareness'' the ``Pre-crash Sensing Warning'' use case has more stringent latency requirements comparing to the rest of the use cases of this category.} & 	V2V/V2I	&100ms-1sec	&90-95\%	&5-96\\
%
%Cooperative Sensing	& V2V/V2I	&10ms-1sec	&$>$95\%	&5-25000\\
%
%%Cooperative Awareness & 	V2V/V2I	&100ms-1sec	&90-95\%	&5-40\\
%
%Cooperative Maneuver 	& V2V/V2I	& $<$10ms-x100ms	&$>$99\%	& 10-5000\\
%
%Vulnerable Road User 	& V2P	&100ms-1sec	& 95\%	&5-10 \\
%
%Traffic Efficiency	&V2N/V2I	& X1sec	&$<$95\%	&10-2000\\
%
%Tele-operated Driving	& V2N	& $<<$100ms	& $>$99\%	&$>$25000\\
%\end{tabular}
%\end{table}

\textbf{Cooperative awareness} use cases are based on periodic messages that are broadcasted with a transmission rate between 1 and 10~Hz and the payload of each message ranges from 60 to 1500 Bytes~\cite{etsi14}, affected by the road and traffic conditions. Due to the dynamic nature of cooperative awareness applications the data rate is between 5~kbps to 40~kbps and the required reliability is 90-95\%~\cite{3gppTR22885}. The end-to-end latency and the required reliability for the \textbf{cooperative sensing} are affected by the triggering event. 
%For instance, for the exchange of raw sensor data in a crash mitigation scenario a large amount of raw data (25Mbps) should be transmitted with very high reliability ($>>$99\%) within 10 ms~\cite{alliance20155g_2}. The 25Mbps data rate is needed in the case that LIDAR (e.g., 64 lasers/detectors, 5-15Hz spin rate) and sensor data are transmitted. In other use cases that a H.265/HEVC High Definition (HD) video stream should be established to increase the situational awareness of vehicles then 10~Mb/s are sent from the source vehicle. 
For instance, for the exchange of raw sensor data in a crash mitigation scenario a large amount of raw data (25~Mbps) should be transmitted with very high reliability ($>$99\%) within 3 ms~\cite{alliance20155g_2,3gppTR22886}. Data rate of 25~Mbps is needed in the case when LIDAR (e.g., 64 lasers/detectors, 5-15~Hz spin rate) and sensor data are transmitted. In other use cases where H.265/HEVC High Definition (HD) video stream should be established to increase the situational awareness of vehicles, 10~Mb/s are sent from the source vehicle.

\textbf{Cooperative maneuvers} (Fig.~\ref{fig:FutureUseCases}) include a wide range of use cases (e.g., cooperative collision avoidance, cooperative lane change) with different latency and reliability requirements. Cooperative collision avoidance requires the exchange of planned trajectories in 10~ms, especially in urban environments with very high reliability ($>$99\%). The maximum expected data rate for each vehicle (1.3~Mbps) is derived from the trajectories (i.e. 12~Bytes/coordinate, 10~ms resolution for a trajectory of 5 seconds) that are transmitted with a periodicity of 10~ms in case of emergency situations such as cooperative collision avoidance. Dense platooning is a special case of the ``Cooperative Maneuver'' category, which can have latency requirement as stringent as 3~ms, reliability higher than 99\% and data rate that higher than 25000 kbps, if sensor sharing is required.

\textbf{Vulnerable road user} use cases have a behavior similar to the cooperative awareness category (i.e., latency and reliability requirements) with the difference that the destination device is a user equipment (e.g., smartphone), where the needed information is less (payload size from 60 to 120 Bytes), which consequently reduces the expected data rate (5-10~kbps). 

\textbf{Traffic efficiency} is supported using the V2N or the V2I modes without strict delay or reliability requirement, since there is no need for prompt (re-)action at the vehicle side. Each vehicle updates the Traffic Management server (uplink) every few seconds with location, status and road information, which are required for the more efficient route selection (the payload of this type of message is 1500 Bytes~\cite{etsi14}). The response from the Traffic Management servers (downlink) includes digital map updates (2~MBytes). 

\textbf{Teleoperated Driving} requires high uplink data rate (up to 25~Mbps) and much lower downlink data rate for the control and command actions that are sent from the operator to the vehicle (e.g., lateral or longitudinal control). The 25~Mbps uplink data rate is derived from two or more cameras (e.g., front, side, back) and other sensor information that is required to provide to the human operator experience similar to that of a regular driver the car. The resulting traffic demand is approximately 10 Mb/s for each H.265/HEVC High Definition (HD) video stream. An end-to-end latency of less than 20~ms for fast vehicle control and feedback is needed together with reliability higher than 99\% to avoid application malfunctions~\cite{kwoczek16,3gppTR22886}.

\begin{table*}[t!]
\centering
\caption{Performance Requirements of Different V2X Use Cases derived from~\cite{3gppTS22186}} 
\label{tab:UseCases}
\begin{tabular}{l}
\begin{tabular}{l c c c c c}
\textbf{Use Case Type}	& \textbf{V2X Mode}	& \textbf{End-to-End Latency}	& \textbf{Reliability}	& \textbf{Data Rate per veh. (kbps)} & \textbf{Comm. Range$^\dag$}\\
Cooperative Awareness%$^\dag$\tablefootnote{In the ``Cooperative Awareness'' the ``Pre-crash Sensing Warning'' use case has more stringent latency requirements comparing to the rest of the use cases of this category.} 
& 	V2V/V2I	&100ms-1sec	&90-95\%	&5-96 & Short to medium\\
Cooperative Sensing	& V2V/V2I	&3ms-1sec	&$>$95\%	&5-25000 & Short\\
Cooperative Maneuver 	& V2V/V2I	& $<$3ms-100ms	&$>$99\%	& 10-5000 & Short to medium\\
Vulnerable Road User 	& V2P	&100ms-1sec	& 95\%	&5-10 & Short\\
Traffic Efficiency	&V2N/V2I	& $>$1sec	&$<$90\%	&10-2000 & Long\\
Teleoperated Driving	& V2N	& 5-20ms	& $>$99\%	&$>$25000 & Long\\
%eV2X use cases here & & & & \\
\end{tabular}\\
\rule{0in}{1.2em}$^\dag$\scriptsize Communication range is qualitatively described as ``short'' for less than 200 meters, ``medium'' from 200 meters to 500 meters, and ``long'' for more than 500 meters.\\   
%\rule{0in}{1.2em}$^\dag$\scriptsize In the ``Cooperative Awareness'' the ``Pre-crash Sensing Warning'' use case has more stringent latency requirements comparing to the rest of the use cases of this category.\\   

 \end{tabular}
\end{table*}

\begin{figure*}[!t]
	\begin{center}
		\includegraphics[trim=2cm 6cm 2.3cm 4.5cm,clip=true,width=0.7\textwidth]{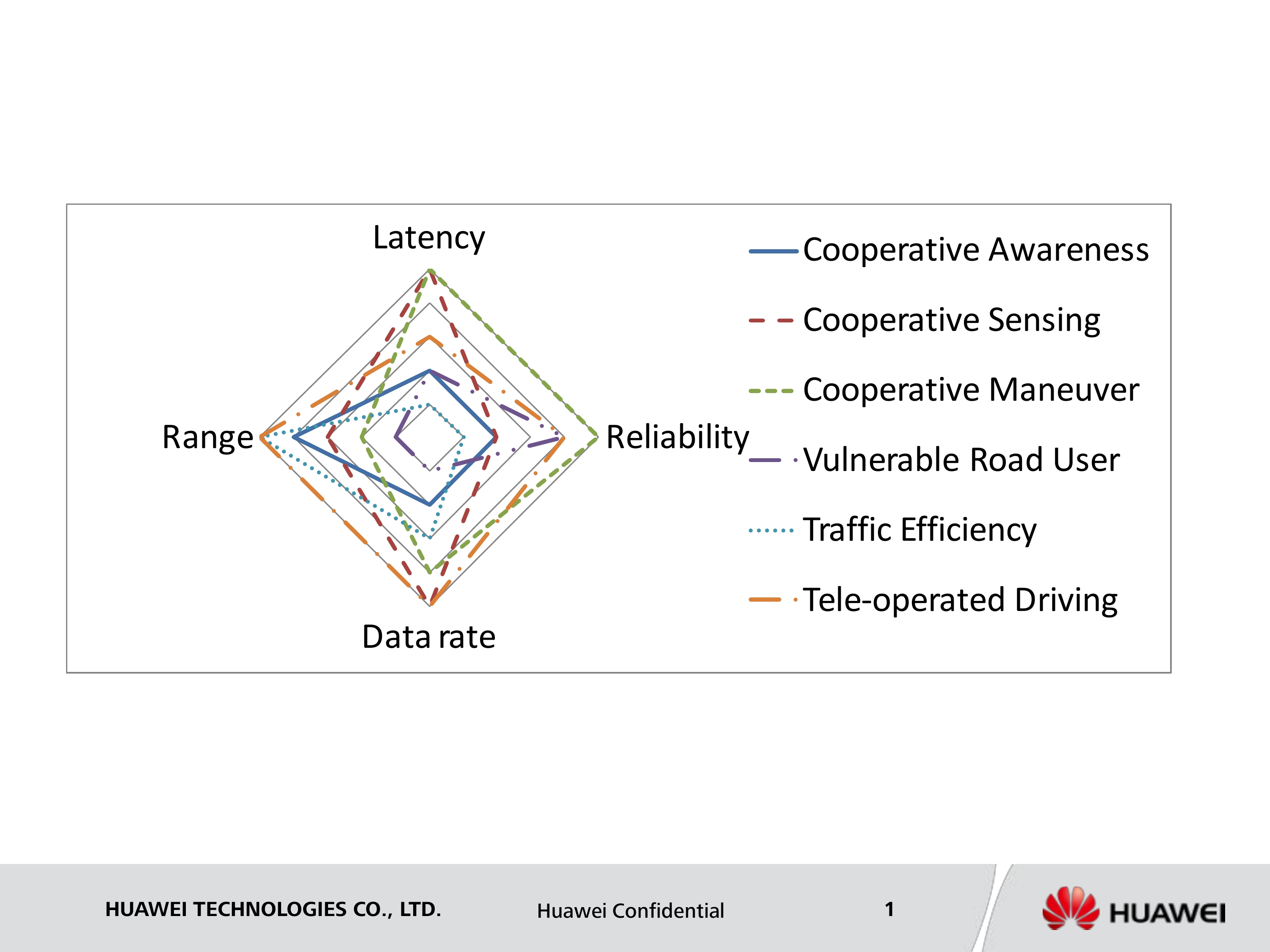}
	\end{center}
	\caption{Use case categories and the stringency of their requirements according to Table~\ref{tab:UseCases}. Higher value represent more stringent requirement (e.g., larger range or lower latency). While neither of the use case categories results in high requirements in all dimensions, combined they ask for a communications system that is able to support long range, low latency, high reliability, and high data rate.}
	\label{fig:spiderGraph1}
\end{figure*}

%% file: commSystems.tex
\section{V2X Communication Systems} % and the road to 5G V2X%and 5G Architectures
\label{sec:CommSystems}

 %Figure~\ref{fig:Arch2} shows the proposed architecture for providing V2X services in a future 5G network. 5G V2X network is composed of several existing communication sub-systems, which need to be integrated into a new 5G architecture to meet the V2X requirements. This section briefly describes the system components and discusses how to further extend and integrate them. %, whereas we defer their suitability for each of the V2X use cases to Section~\ref{sec:UseCases}.

This section describes existing and future communication systems, which are envisioned as components of a 5G network providing V2X services and will need to be integrated into a new 5G architecture, as shown in Figure~\ref{fig:Arch2}. In what follows, cellular network in cm- and mm-Wave frequency bands, IEEE 802.11p and Vehicular Visible Light Communication are briefly discussed.

\subsection{Cellular centimeter-wave}
3GPP has recently finalized the first set of cellular standards for V2X communication, under the name LTE-V2X~\cite{seo2016lte}. 
%Elaborating on the proposal described in~\cite{seo2016lte}, 
Building on top of the existing LTE-V2X standards, future 5G V2X cellular systems will provide the main radio interface to support V2X communication through three distinct paradigms shown in Fig.~\ref{fig:Arch2}: i) Cellular V2X; ii) Cellular-assisted V2V; and iii) Cellular-unassisted V2V. %In all cases, we assume that user equipment (UE) is embedded in the vehicle, with appropriate interconnection with vehicle's onboard systems and externally mounted antennas.
%3GPP has recently finalized the specification of the first set of cellular standards for V2X communication, under the name LTE-V2X~\cite{seo2016lte}. 
In the context of cellular networks, V2V communication is also referred to as sidelink when discussing direct V2V communication, either in assisted or unassisted mode. Furthermore, V2V communication can also be enabled indirectly wherein the base station (and possibly other network elements) relay the information between vehicles. In the LTE-V2X context, direct sidelink communication (be it assisted or unassisted) is enabled through the PC5 interface. The terms uplink and downlink describe communication between a vehicle and a base station, irrespective of whether the base station is the communication counterpart for the vehicle (in case of V2I) or not (in case of V2N). In the LTE-V2X context, uplink/downlink communication is enabled through the Uu interface.

\emph{Cellular V2X} refers to ``classic'' cellular uplink/downlink communication, wherein a vehicle equipped with a UE is communicating with a base station to reach either an edge server (e.g., to update the real-time maps for the road ahead), backend server (e.g., to obtain traffic update from traffic center), or to connect to another (possibly far away) vehicle. Apart from macro cells, this also includes small cells, for example in the form of Roadside Units (RSUs), as shown in the bottom left of Fig.~\ref{fig:Arch2}. In addition to improving coverage, RSUs increase the overall cell throughput, and may further allow them having enhanced functionalities % compared to IEEE 802.11p stations. Such could include 
such as fast radio access and handover between RSUs and coordinated resource allocation for reducing the latency. %Communication in this Vehicle-to-Infrastructure (V2I) sense can also serve the purpose of obtaining local information, e.g. on signal phase and timing if the base station is connected to a traffic light or on hazardous conditions on the road ahead.

\emph{Cellular-assisted V2V} %(also referred to as ``operator-assisted'' in~\cite{khelil2014suitability}), 
is a scheme where the base station coordinates the communication between vehicles by providing control information incl. link scheduling and resource allocation instructions to vehicles,
%instructing them which resources to use and for how long (e.g., in terms of time or frequency allocation), which vehicles to communicate with (e.g., managing priority in case of different V2V applications competing for resources~\cite{3gppTR22885}).
or manage priorities in case of different V2V applications competing for resources~\cite{3gppTR22885}. This mode of communication is best suited for extremely low latency and high reliability V2V communication, as the base station ensures that resources will be available when needed and time-consuming data transmission over the cellular network is avoided. In addition to enabling low latency, cellular V2V can also be used for traffic offloading. For example, the data that is relevant to a small geographical region only (e.g., region surrounding the transmitting vehicles) can be transmitted via V2V, thus relieving the pressure on the core network~\cite{3gppTS23401}.

\emph{Cellular-unassisted V2V} %, also called operator-free~\cite{khelil2014suitability}, shall be used for direct Ad Hoc 
is a mode where vehicles communicate directly, without assistance from the base station. It is similar to IEEE 802.11p in the sense that it does not involve a base station, and is therefore useful especially for information exchange in out-of-coverage scenarios. However, cellular-unassisted V2V has some additional features compared to IEEE 802.11p. First, it can reuse part of the cellular frequency band in areas without or with partial cellular coverage for communication between vehicles. Resource usage can be considered to be under tacit control of the cellular network (i.e., the network determines which resources vehicles can use), which also simplifies a re-transition back to cellular V2X. Moreover, out-of-coverage users can remain time-synchronized to the cellular network in order to communicate with in-coverage users, enable interference-free multi-link connectivity and quick re-attachment to a base station if back in coverage. In this sense, even when they are out-of-coverage or in partial coverage, vehicles can be considered as part of the cellular network.\\

 %In case of non-safety applications, it is also conceivable to utilize the unlicensed spectrum (e.g., ISM bands).  

%The existing V2V connection can further assist users in accessing the serving base station, and maintain continuity in resource usage. In case of no cellular coverage, resources are likely to be available for at least a subset of communicating vehicles, since by definition there is no interference between vehicles and base station\footnote{\konstantinos{The fact that there is no base station to interfere with does not mean that resources are available without any constraint for out of coverage V2V.}}. 

\subsection{Millimeter-wave communication for V2X}

%\mate{Add Heath + Toyota paper here} 
%Riding on the trend of reduced cell sizes 
In search for increased network capacity through reduced cell sizes and for more available spectrum, millimeter wave (mmWave) band has seen a resurgent research interest. The vision we have for the role of mmWave in V2X is twofold:
\begin{itemize}
	\item V2I communication, wherein short-lived, high data rate connection can be established between vehicle and nearby base stations (small cells) to exchange delay-insensitive data (e.g., map updates and infotainment data in the downlink, and collected traffic and sensor information for large scale traffic monitoring in the uplink).
	\item Directional V2V communication for supporting particular use cases, such as back-to-front (bumper-to-bumper) communication between subsequent vehicles in a platoon~\cite{Va2016}.
\end{itemize}

Although mmWave communication is very attractive from the data throughput perspective, it brings challenges on the physical layer. Due to high propagation path losses and its susceptibility to shadowing, it is deemed suitable for mostly short range (a few hundred meters) and point-to-point LOS communication~\cite{rangan2014millimeter}. Furthermore, since the Doppler spread is linear in the carrier frequency, in the 60~GHz band it will be ten to thirty times that of in 2-6~GHz band. In order to deal with these effects, a special frame design and numerology will be needed along with adjusted solutions, still target on mmWave communication for low (relative) mobilities as in the scenarios mentioned above. % thus limiting the mmWave communication to lower speed urban environments and highly directional, same way highway communication.

Radio signals used for communication with the cellular network and between vehicles can be used for position estimation. The achievable positioning accuracy can be significantly higher for 5G networks that for legacy LTE networks, due to the higher signal bandwidth and the dense nature of the network, which naturally enables LOS with a high probability. Estimation of the relative position between vehicles is directly useful for particular use cases as platooning, but can also serve as input to the cellular network in order to improve the estimation of vehicles' absolute position.

\subsection{IEEE 802.11p}
Radios implementing the IEEE 802.11p protocol suite -- based on IEEE 802.11a WiFi standard -- are expected to support V2X communication in the U.S. (under the guise of Wireless Access in Vehicular Environments -- WAVE -- specified in the IEEE 1609 protocol set), Europe (under the name ITS-G5 and specified by ETSI TC ITS~\cite{etsi202663}), as well as elsewhere. The incentive for using the technology came as U.S. and Europe reserved 75~MHz and 30~MHz of licensed band respectively in the 5.9~GHz frequency range. Recently, there has been discussions of a possible mandate of the technology by U.S. regulatory bodies. Furthermore, some vehicle manufacturers (most notably, GM in the U.S., and Toyota in Japan) announced proactive deployment of V2X systems based on IEEE 802.11p radios in the year 2016. The most prominent facilities provided by IEEE 802.11p-based V2X are cooperative awareness functionality (e.g., in Europe, through Cooperative Awareness Messages -- CAMs~\cite{ETSIEN3026372}) and event-based emergency updates for hazardous situations (e.g., through Decentralized Environmental Notification Message -- DENM~\cite{ETSIEN3026373}). From the technical perspective, IEEE 802.11p physical layer is well-designed for V2V communication, as it is robust to Doppler spread, and can also enable low-latency communication with its short radio frame. However, it suffers from a high collision probability under medium or high traffic loads due to its simple random access scheme, making it not very suitable for ultra-reliable V2V communication.

\subsection{Vehicular Visible Light Communication}
Vehicular Visible Light Communication (VVLC) has the capability to enable illumination/signaling, communication, and positioning, in a single communications system~\cite{yu2013smart}. VVLC assumes that the light emitting diodes (LEDs) in vehicle headlights and taillights are used to transmit information, whereas photodiodes or cameras serve as receivers. Since LED lights, cameras, and photo diodes are becoming commonplace on vehicles, the cost of adding VVLC capability to the cars is limited to inexpensive electric circuitry that enables communication. VVLC also has centimeter-grade relative positioning capability, due to the dual headlight/taillight configuration, and the known dimensions of vehicles~\cite{roberts2010visible}. It has the characteristics making it particularly suitable for precise outdoor (road) and indoor (parking) positioning, low cost installations (e.g., on scooters, which cannot subsume complex communications system in their price), and as support for RF V2X communication (i.e., enabling fault tolerant behavior through multi-communication sensor fusion). The positioning capabilities in particular make VVLC a useful complement to satellite-based positioning systems (which have meter-grade precision); the two technologies can be combined to enable some of the key V2X use cases (e.g., cooperative lane merging and platooning~\cite{3gppTR22885}). %and a low-cost alternative to radar-based positioning (with the advantage of communication ability). %(i.e., vehicles can use VVLC to communicate their trajectories 
Main performance characteristics of VVLC communication include a range of at least 50~m (with low BER) and a bitrate up to 2~Mb/s @ 50~m and 500+ Mb/s @ 5~m \cite{yu2013smart}. VVLC further offers directional communication with a high LOS probability, which results in limited interference and a high reuse factor. Finally, due to separated transmitter (LED) and receiver (photo diode/camera), full duplex is feasible in VVLC communication.

\section{Gap Analysis and %System design 
Architectural Aspects Towards 5G V2X}\label{sec:Gap} %on the road to 5G V2X}

This section discusses the performance of existing communication technologies such as LTE-V2X~\cite{seo2016lte} and IEEE 802.11p as well as oncoming mmWave and VVLC, as these are described in Section~\ref{sec:CommSystems}, and identifies the gap between their performance and the requirements of 5G use cases. Table~\ref{tab:Mapping} provides a qualitative evaluation of the different communication technologies capability to fulfill the requirements of categorized use cases, based on the information from Table~\ref{tab:UseCases}. The extent of fulfilling the requirements (latency, reliability, data rate and coverage) of each use case leads to individual evaluation results.
\begin{table*}[t!]
	\centering
	\caption{Qualitative Assessment of the Ability of Communication Technology to Support Use Cases} 
	\label{tab:Mapping}
	\begin{small}
		Suitability legend: \\ ``\checkmark\checkmark'': suitable technology to support the use case and requirements under all circumstances with no (or with minor) configuration; \\ ``\checkmark'': suitable technology to support the use case and performance requirements under specific conditions (e.g., low congestion level); \\``-'': not suitable technology because the specific use case or its performance requirements are not supported\\[+5pt]
	\end{small}
	\begin{tabular}{l c c c c}
		\textbf{Use Case Type}	& \textbf{LTE-V2X}	& \textbf{802.11p} 	&  \textbf{mmWave}	& \textbf{VVLC}\\ \hline
		\textbf{Cooperative Awareness} &&&& \\
		Emergency Vehicle Warning~\cite{etsi14}	&\checkmark\checkmark	&\checkmark\checkmark	&-	&-\\ 
		Forward Collision Warning~\cite{etsi14} &\checkmark\checkmark	&\checkmark\checkmark	&\checkmark	&\checkmark\\ \hline
		
		\textbf{Cooperative Sensing} &&&& \\
		See-through~\cite{alliance20155g_2}	&\checkmark	&\checkmark	&\checkmark\checkmark	&\checkmark\\
		Sensor Sharing~\cite{3gppTR22886}	&\checkmark	&\checkmark	&\checkmark	&\checkmark\\ \hline
		
		\textbf{Cooperative Maneuver} &&&& \\
		Platooning~\cite{3gppTR22885}	&\checkmark\checkmark	&\checkmark	&\checkmark	&\checkmark \\
		High Density Platooning~\cite{3gppTR22886}	&-	&-	&-	&- \\
		Cooperative Adaptive Cruise Control~\cite{3gppTR22886}	&\checkmark	&\checkmark	&-	&-\\ \hline
		Cooperative Intersection Control~\cite{alliance20155g_2}	&\checkmark	&\checkmark	&-	&- \\ \hline
		
		\textbf{Vulnerable Road User}~\cite{etsi14}	&\checkmark	&\checkmark	&-	&-\\ \hline
		\textbf{Traffic Efficiency}~\cite{etsi14}	&\checkmark\checkmark &\checkmark	&-	&-\\ \hline
		\textbf{Tele-operated Driving}~\cite{3gppTR22886}	&\checkmark	&-	&-	&-\\ \hline
		
	\end{tabular}
\end{table*}

\begin{figure*}[!t]
	\begin{center}
		\includegraphics[trim=5cm 5.9cm 5cm 5.8cm,clip=true,width=0.7\textwidth]{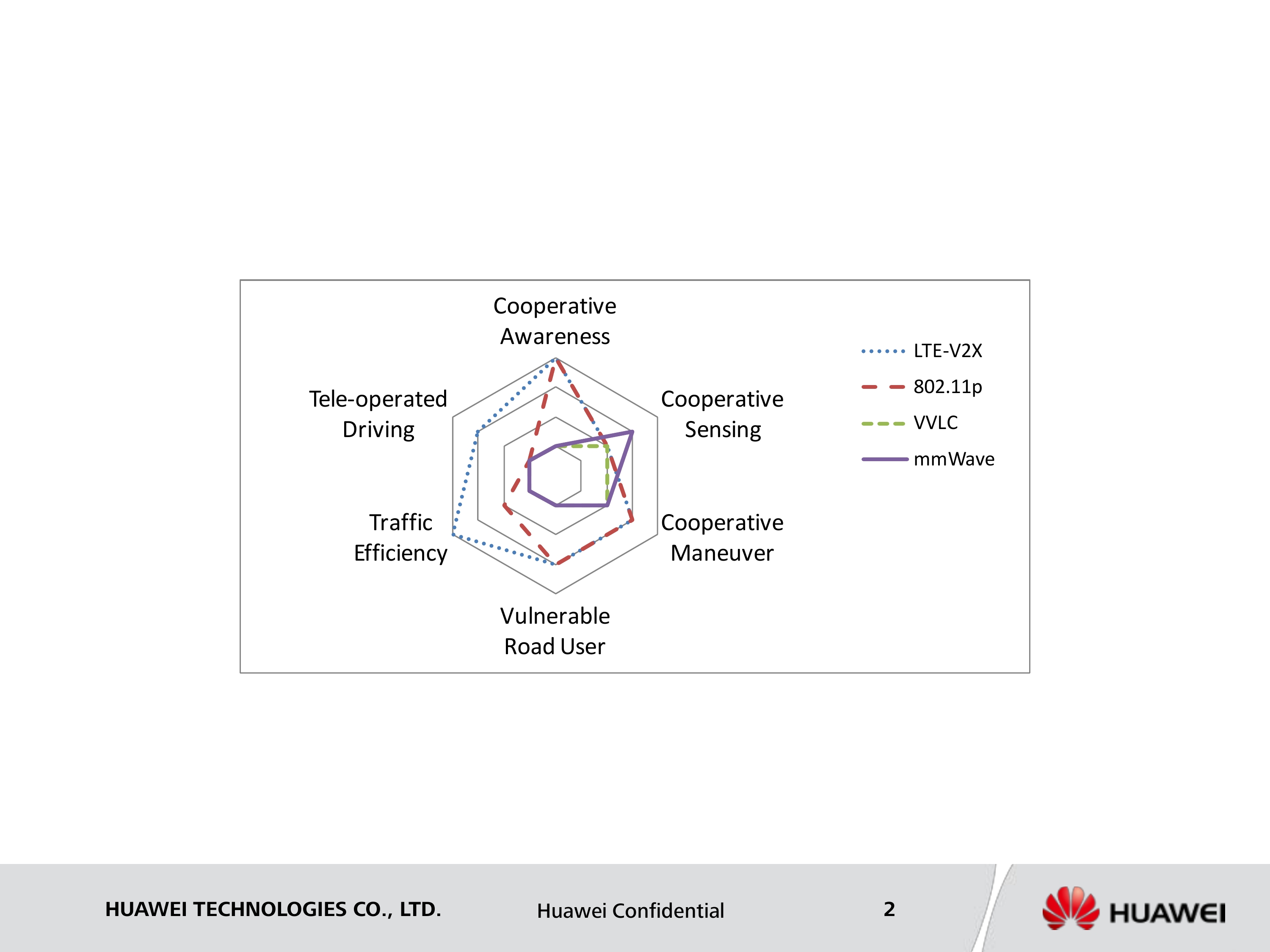}
	\end{center}
	\caption{Ability of communication technologies to support categories of use case category, based on Table~\ref{tab:Mapping}. The graph indicates that neither of the existing technologies can enable all use case categories.}
	\label{fig:spiderGraph2}
\end{figure*}

As can be seen from Fig.~\ref{fig:spiderGraph2}, %observed from Table~\ref{tab:Mapping}, 
both LTE-based V2X and 802.11p are suitable technologies to support cooperative awareness, cooperative sensing, cooperative maneuvers and VRU use cases, but only under certain conditions. The performance of IEEE 802.11p is degraded in high load, dense network environments~\cite{mir2014lte}, which means that the use cases such as platooning can be supported in low load scenarios, while in high load scenarios the delay and reliability requirements cannot be guaranteed. On the other hand, LTE-based system, particularly in cellular assisted mode with more efficient resource management, can better support platooning. However, the existing functions and procedures of the LTE-based system have not been designed to meet the very tight delay and reliability requirements, especially for use cases such as high density platooning and other advanced Cooperative Sensing and Cooperative Maneuver use cases (Table~\ref{tab:UseCases}). Some of the use cases that belong to the Cooperative Awareness category (e.g., Emergency Vehicle Warning) do not have strict delay and reliability requirements (i.e., 100ms-1sec delay, 90-95\% reliability) and could be supported by both technologies under all circumstances. The use cases that require communication with a network entity (e.g., Tele-operated Driving, Traffic Efficiency) can only be served by cellular-based technologies. However, the most stringent delay and reliability requirements (e.g., for the remote control of a vehicle) cannot be met with the current LTE-based cellular system; therefore, next generation cellular systems need to be considered for these use cases. 

On the other hand, VVLC and mmWave are mainly suitable for use cases such as see-through and platooning (for direct communication between a leading and a following vehicle), where communication occurs over short range links in line-of-sight conditions. However, these technologies cannot provide the required data rate and reliability requirements if the vehicles are in non-line-of-sight (NLOS) or the required range is large. Inability to support stringent requirements in NLOS and at large distances is why VVLC and mmWave cannot support the advanced use cases as stand-alone technologies. One solution to this problem is to use VVLC and mmWave in combination with other technologies operating in lower frequencies, where they can provide high data rate, low interference support; for example, in case of high density platooning, cmWave cellular system can be used for coordination and be supplemented by mmWave for communication between members of platoon.

Fig.~\ref{fig:spiderGraph2} visualizes the ability of communication technologies to support use case categories in form of a spider graph. While these communication systems can support use cases that have less stringent requirements (e.g., Traffic Efficiency by LTE-V2X and Cooperative Awareness by both LTE-V2X and IEEE 802.11p), any one technology falls short of supporting the complete set of requirements. In fact, serving all envisioned 5G use cases would probably not even be possible by a simple superposition of the above technologies. %However, it is interesting that some use cases can be sufficiently served by a single technology, e.g. Traffic Efficiency by LTE-V2X and Cooperative Awareness by both LTE-V2X and IEEE 802.11p. %For these cases, a direct mapping of applications to communication systems is possible and also recommended.

\begin{figure*}[!t]
	\begin{center}
		\includegraphics[trim=0cm 3cm 0cm 0cm,clip=true,width=0.95\textwidth]{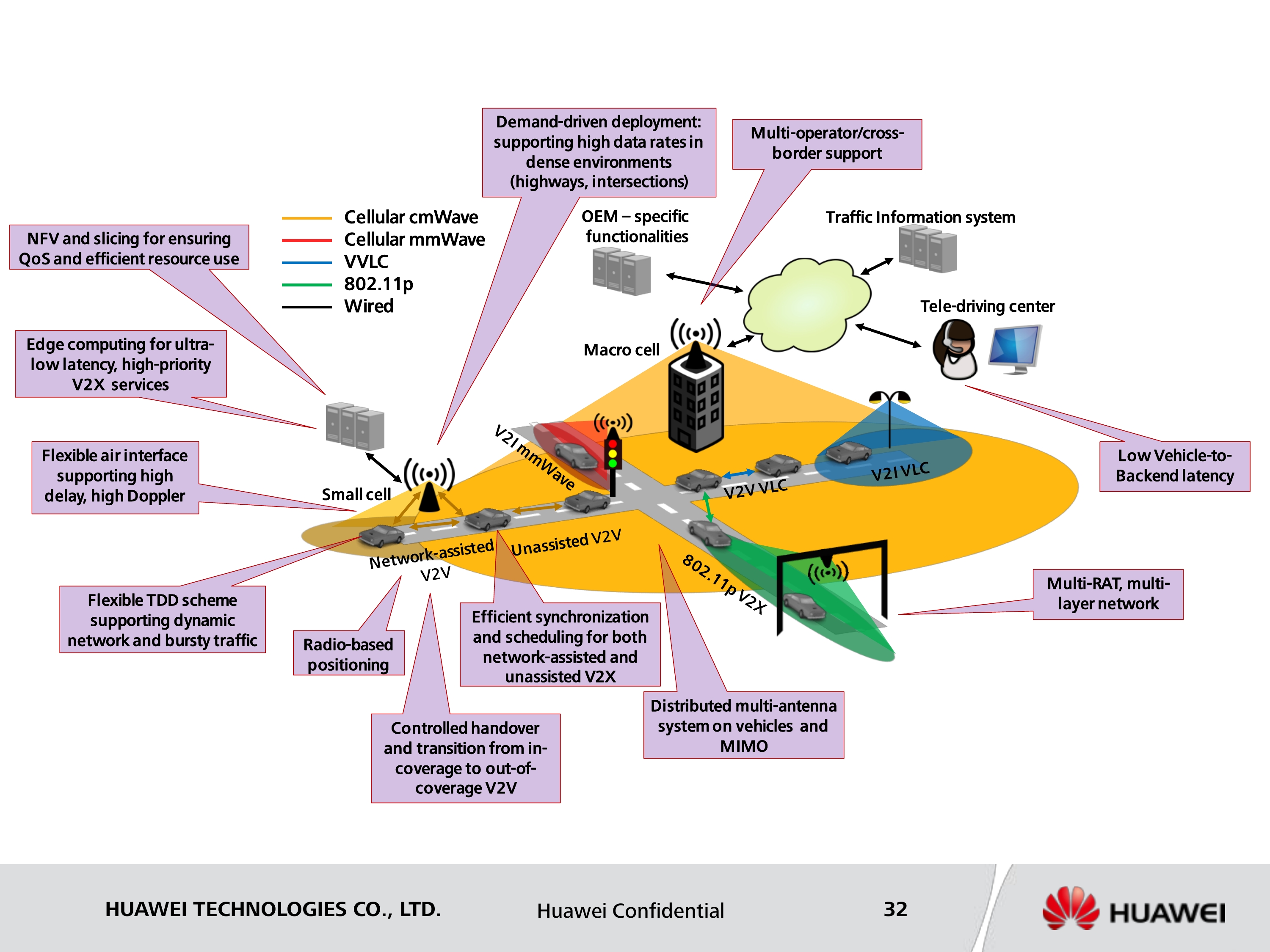}
	\end{center}
	\caption{Heterogeneous Multi-RAT network architecture detailing components of 5G V2X for 5G V2X required to enable all use case categories, eventually leading to accident-free connected automated driving.
	 %\mate{Fig has been updated for the 2nd paper. Should we use that updated version in this paper as well?}
	}
	\label{fig:Arch2}
\end{figure*}

In what follows, we discuss how existing and future communication systems can be leveraged as part of a 5G multi-RAT solution. %	Using the motivating example from Sect.~\ref{sec:Motivation} and 
Based on the requirements described in Sect.~\ref{sec:UseCases} and the gap analysis discussion, Fig.~\ref{fig:Arch2} depicts the components of a 5G V2X Radio Access Network (RAN) architecture. As shown, a new cmWave macro-cellular system will come into play, which may in a first phase coexist with LTE-based cellular system, as well as with IEEE 802.11p. The role of the macro-cellular network will be to provide large coverage, high data rates, low latency for data and control information. It will also connect with RSUs, small cells and other infrastructure units.

Due to their beneficial characteristics in terms of increased bandwidth, strong directionality/low interference, and ability to support network densification, we expect technologies such as mmWave and VVLC to be incorporated in the 5G V2X access network architecture to support specific V2X use cases (see Fig.~\ref{fig:spiderGraph2}). %Table~\ref{tab:Mapping}). 
These technologies will provide short-range high-throughput communication as part of an advanced 5G V2X system. The coexistence with current systems and gradual inclusion of new radio interfaces and technologies brings up the important design goal for 5G radio access architecture: the capability to support diverse technologies and enable backward and forward compatibility.  

Fig.~\ref{fig:Arch2} depicts further technologies not discussed in this paper, which nevertheless need to be in place as part of the 5G V2X solution. These range from radio-based positioning, which can be enabled more efficiently thanks to increased bandwidth in mmWave frequencies, to more advanced scheduling, which can incorporate geographic information for better decision making, to Mobile Edge Computing (MEC)~\cite{patel2014mobile}, which
  can provide cloud-computing and networking capabilities within the Radio Access Network in close proximity to vehicles. Some of the physical layer aspects shown in Fig.~\ref{fig:Arch2} (e.g., flexible air interface, MIMO, and synchronization) have been discussed in~\cite{boban2016design}. All of these technologies require further study and innovative designs in order to enable the stringent 5G V2X use cases.

\iffalse 

%\subsubsection{System Integration}
Due to the reservation of bandwidth in the 5.9~GHz band, the likely government mandates on the use of the technology, and steady research and development over the past decade, IEEE 802.11p-based systems %operating in 5.9~GHz band 
will likely be deployed in some markets as the initial V2X technology. Similarly, because telecom operators and vendors have worked on previous iteration of the system for decades, an LTE-based cellular system (as described in~\cite{seo2016lte}) %and~\cite{sun2016LTE}) 
will be the initial step towards a new 5G V2X communication system. Therefore, alongside IEEE 802.11p and an LTE-based cellular network, it is expected that early stage of 5G V2X will be operating in 2-6~GHz bands %. % and IEEE 802.11p-based systems operating in 5.9~GHz band.  
and providing the aforementioned cooperative awareness and emergency messages along with high-throughput cellular communication. %``awareness'' capabilities through Cooperative Awareness Message (CAM)~\cite{etsi14} and  event-based emergency updated through Decentralized Environmental Notification Message (DENM).

On the other hand, due to the better capability to support low latency, high reliability, and high throughput requirements, the future 5G systems will be the candidate for supporting use cases leading to automated transport including tele-operation, connected intersection control, cooperative lane merging, etc.
\fi

%% file: conclusions.tex
\section{Conclusions and Outlook}\label{sec:Conclusions}
%\section{Connected Automated Car: the cheapest way of doubling road and parking capacity}
%Connected automated vehicles are not only a convenient mode of transportation, but also the cheapest way of doubling road and parking capacity. 
%\subsection{Benefits of traffic coordinated through communication}
Benefits of  vehicles cooperating through V2X communication 
are numerous: i) they can enable safer journey; ii) % reduce the inter-vehicle spacing, 
improve highway, intersection, and parking lot capacity; and iii) enable seamless end-to-end mobility.
%Communicating vehicles can have the following benefits over non-communicating vehicles:
%\begin{itemize}
%\item Reduce the inter-vehicle spacing (Ref)
%\item Optimize the intersection throughput (Ref)
%\end{itemize}
%The above improvements apply even when compared to fully automated, non-communicating vehicles: the speed feedback on the dynamics of other vehicles received through communication cannot be matched by sensors on the vehicles; for example, irrespective of how reactive the radar or LIDAR is, due to physical constraints, it will still take some time until it detects sudden breaking of the vehicle in front. 
Furthermore, in non-line-of-sight (``behind-the-corner'') conditions, V2X communication is the only technology that can enable safe behavior through almost instantaneous exchange of vehicle dynamics.

This paper has described the relevant use cases and analyzed their requirements, which the future V2X communications systems needs to support. The most demanding use cases require high link reliability (in come cases, above 99\%), low latency (below 10~ms), and high throughput (tens of Mb/s per vehicle), often concurrently (Fig.~\ref{fig:spiderGraph1}). 
We also performed a qualitative gap analysis of the capability of existing technologies (Fig.~\ref{fig:spiderGraph1}) and concluded that the stringent requirements of some use cases cannot be supported by any currently available technology. In order to bridge the gap, we have laid out the design considerations for the 5G V2X system required to 
 %In this paper, we discussed the important remaining aspects t
%Discussions and results presented in this paper are the first step in 
enable next-generation V2X use cases~\cite{3gppTR22886}. 

Significant work remains ahead, ranging from physical layer (e.g., further enhancements of MIMO, fast and reliable coding schemes, initial access and synchronization, frame structure, channel modeling, etc.) to inter-operation  and coordination of multi-RAT systems, to cross-network resource reservation through Network functions virtualization (NFV). Our analysis has shown that connected transportation is one of the most stringent verticals to be supported. In other words, 5G V2X needs to be at the center-point of 5G new radio development, as it requires low-latency, high-reliability, and potentially high-throughput network to ensure accident-free, efficient future transportation.

%\mate{Not sure the part below connects to above. Konstantinos, Apostolos: care to take a go at finishing conclusions?}
To that end, the ongoing work in different fora is aiming to enable ultra low latency, high reliability communication for the future 5G system in general, including support for different vertical industries. More recently, the newly formed 5G Automotive Association~\footnote{\url{http://5gaa.org/}} has focused specifically on 5G V2X, with strong participation of automotive industry. The activities in 5GAA range from prioritizing existing use cases~\cite{etsi14,3gppTR22886} and defining new ones, to fine-tuning the key performance requirements~\cite{3gppTS22186}, to laying the groundwork for new V2X architecture. These activities are intended to accelerate the work on 5G V2X and support the standardization bodies (3GPP in particular) in designing the communication system that meets the stringent needs of 5G V2X use cases.

%% file: ack.tex
\section*{Acknowledgements}
We would like to thank Liang Hu, Serkan Ayaz, Daniel Medina and Chan Zhou for useful comments and suggestions that improved the quality of the manuscript.